\documentstyle[12pt,bezier]{article}
\def\beq{\begin{equation}}
\def\eeq{\end{equation}}

\begin{document}
\begin{titlepage}
\begin{center}
{\Large \bf Theoretical Physics Institute \\
University of Minnesota \\}  \end{center}
\vspace{0.3in}
\begin{flushright}
TPI-MINN-93/01-T \\
January 1993
\end{flushright}
\vspace{0.4in}
\begin{center}
{\Large \bf Calculation of processes involving many particles at
the kinematical threshold\\}
\vspace{0.2in} {\bf B.H. Smith  \\ }
School of Physics and Astronomy \\
University of Minnesota \\
Minneapolis, MN 55455 \\

\vspace{0.2in}
{\bf   Abstract  \\ }
\end{center}
A diagrammatic technique is derived for calculating processes involving
the production of large numbers of particles. As an example,
the amplitude of three particles scattering into $n$ particles at the
kinematical threshold in $\lambda\phi^4$ theory is calculated. From the
$3 \to n$ amplitude, it is seen that the one-loop amplitude
for $2 \to n$ processes is suppressed by a factor of $1/n$.

%PACS numbers: 11.10.Jj, 11.90.+t
\end{titlepage}

It has been shown that one can study
multiscalar production at the tree-level by using recursion
relations$^{\cite{voloshin1,akp1}}$ or the reduction formula
method$^{\cite{brown}}$. These techniques enable one to calculate the
tree-level amplitude of one off-shell scalar particle decaying into
many scalar particles at rest.

These methods have been shown to be applicable to other problems. The
one-loop correction was found in $\lambda\phi^4$ theory for both the case of
unbroken$^{\cite{voloshin3}}$ and broken$^{\cite{smith}}$ reflection
symmetry. In finding these corrections the amplitude of $2$ on-shell
scalar particles producing $n$ particles at the kinematical threshold
was discovered to vanish for $n>4$ in the case of unbroken symmetry and
$n>2$ in the case of broken symmetry (for a discussion on this
phenomena see $\cite{voloshin4}$). This effect was further discussed by
Argyres, Kleiss, and Papadopoulos$^{\cite{akp2}}$ and by Brown and
Zhai$^{\cite{bz}}$. Argyres, Kleiss, and Papadopolous have discussed
the cross channel of this process (i.e. one off-shell particle decaying
into $n$ particles at rest and one particle above the kinematic threshold).

In this paper, we discuss a diagrammatic technique for calculating
these amplitudes. We show how to build a set of Feynman rules for
the generating functions of processes involving multiparticle production.
With these rules, one can consider processes involving $M$
scalars scattering into $N$ scalars with none of the $M$ or $N$ particles
neccesarily being at rest. However, instead of allowing one to calculate
the $M \to N$ amplitude as in normal diagrammatic techniques, this
technique allows one to calculate the generating function for $M \to N + n$
processes where the $n$ particles are at rest.

It must be emphasized that this procedure does not allow the study of
many particles above the kinematical threshold. While large numbers of
particles at rest may be produced, the calculations become more and
more complicated as the number of particles above the threshold is
increased. This approach does provide a compact method to
calculate the amplitude for a few particles scattering to produce a few
particles above the threshold and many particles at the kinematical
threshold.

We shall demonstrate this technique by calculating the $3 \to n$
amplitude in $\lambda\phi^4$ theory. Through the unitarity relations,
this amplitude provides an easy way to study the one-loop properties
of $2 \to n$ processes.

We will be discussing $\lambda \phi^4$ theory with the Lagrangian,
\beq
{\cal L} = {1\over{2}}(\partial\phi)^2 - {1\over{2}}m^2\phi^2 -
{\lambda\over{4}}\phi^4.
\label{langrange}
\eeq
We will primarily be deriving Feynman rules for the case of unbroken
reflection symmetry ($m^{2}>0$). Most of what is discussed may be trivially
extended to the case of broken symmetry as well as theories
dicussed in $\cite{voloshin5}$, $\cite{smith2}$ and $\cite{bz}$.

We wish to consider the amplitude for an interaction involving $N$ particles
with arbitrary four-momenta. During this interaction, $n$ particles will
be produced at the kinematic threshold. The
particles produced at the threshold will have a
four-momentum denoted as $q \equiv (1,{\bf{0}})$. The generating functions
for these amplitudes can be
denoted as the N-point Green function, $G_{n}^{(N)}(p_1,...,p_n)$. To
calculate this function, it is easiest to consider the amplitude of
processes involving one particle decaying into $N-1$ particles above the
threshold and $n$ particles at the threshold. This amplitude is written as
\beq
A_{n}^{(N)}(p_1,...,p_n) \equiv
G_{n}^{(N+1)}(-P-nq,p_1,...,p_n),
\eeq
where $P \equiv p_1+...+p_n$.

The amplitude $A_{n}^{(N)}$ can be considered recursively as in Fig. 1.
We can write the recursion relation
\begin{eqnarray}
A_{n}^{(N)}(p_1,...,p_n) = {\lambda\over{(P+nq)^2-1}}
\sum_{n_1,n_2,n_3} \delta_{n,n_1+n_2+n_3} \sum_{N_1,N_2,N_3}
\delta_{N,N_1+N_2+N_3}
\nonumber \\
{{n!\over{n_{1}!n_{2}!n_{3}!}} \sum_{perms.}
A_{n_1}^{(N_1)}(\{P_{n_1}\})A_{n_2}^{(N_2)}(\{P_{n_2}
\})A_{n_3}^{(N_3)}}(\{P_{n_3} \}),
\label{recurse}
\end{eqnarray}
where $\{P_{m}\}$ is a set of $m$ of the original momenta and the
second sum runs over all possible permutations of the momenta.

Equation ($\ref{recurse}$) can be simplified by using the function,
\beq
a^{(N)}_{n}(p_1,...,p_n) \equiv -i\left ( {8\over{\lambda}} \right
)^{n/2} {A^{(N)}_{n}(p_1,...,p_n)\over{n!}}.
\label{adef}
\eeq
With this substitution, we can construct a differential
equation for the generating function$^{\cite{akp1}}$,
\beq
g^{(N)}(p_1,...,p_N;x) \equiv \sum_{n=1}^{\infty}
a^{(N)}_{n}(p_1,...,p_N) x^{n}.
\eeq
It can be verified that for all $N>1$, the generating function must
satisfy,
\begin{eqnarray}
\left [ x^{2} {d^{2}\over{dx^2}} + (2P_{0}+1)x {d\over{dx}} + P^{2} - 1
\right ] g^{(N)}(x) = \nonumber \\
\lambda\sum_{N_{1},N_{2},N_{3}} \delta_{N,N_1+N_2+N_3}
g^{(N_1)}(x) g^{(N_2)}(x) g^{(N_3)}(x),
\label{gendiff}
\end{eqnarray}
where the momentum arguments have been omitted for
simplicity and the sum is implied over all permutations of the momenta.

The homogeneous part of equation ($\ref{gendiff}$) contains $g^{(0)}$. This
coefficient is the generating function for multiparticle
production at the threshold, i.e. the amplitude for $1 \to n$ processes
as found in previous works$^{\cite{voloshin1,akp1}}$. Using the fact
that
\beq
\langle n|\phi(0)|0\rangle = (2k+1)!\left ( {\lambda\over{8}} \right
)^{(n-1)/2},
\eeq
$g^{(0)}$ is found to be,
\beq
g^{(0)} = \sqrt{8\over{\lambda}}{x^{2}\over{(1-x^2)^2}}.
\label{gdef}
\eeq
To separate the homogeneous part of equation ($\ref{gendiff}$), one must
simply extract the terms in the sum with  $N_{1}$, $N_{2}$, or
$N_{3}=N$. Equation ($\ref{gendiff}$) can now be recast in the form of the
homogeneous differential equation,
\begin{eqnarray}
\left [ x^{2} {d^{2}\over{dx^2}} + (2P_{0}+1)x
{d\over{dx}} + P^{2} - 1 - {24 x^{2}\over{(1-x^2)^2}}
\right ] g^{(N)}(x) = \nonumber \\
\lambda\sum_{N_{1},N_{2},N_{3}<N} \delta_{N,N_1+N_2+N_3}
g^{(N_{1})}(x) g^{(N_{2})}(x) g^{(N_{3})}(x).
\label{gendiff2}
\end{eqnarray}

As was noted by Voloshin in $\cite{voloshin4}$, the operator on the
left hand side of equation ($\ref{gendiff2}$) can be restated in a familiar
form by making the substitutions,
\begin{eqnarray}
x \to ie^{t} \nonumber \\
g^{(N)}(p_{1},...,p_{N};x) \to e^{-P_{0} t} y^{N}(p_{1},...,p_{N};t).
\label{tsub}
\end{eqnarray}
This substitution transforms the homogeneous operator in equation
($\ref{gendiff2}$) into the Schr\"odinger operator for an exactly
solvable potential. The generating function can now be written as the
solution of
\begin{eqnarray}
\left [ {d^{2}\over{dt^2}}-{\bf{P}}^2-1+{6\over{\cosh^{2}(t)}} \right ]
y^N(t) =  \\ \nonumber
\lambda\sum_{N_{1},N_{2},N_{3}<N} \delta_{N,N_1+N_2+N_3}
y^{N_1}(t) y^{N_2}(t) y^{N_3}(t),
\label{xformedGenDiff}
\end{eqnarray}
where ${\bf{P}}^2$ is square of the magnitude of the spatial part of $P$.

The operator in equation ($\ref{xformedGenDiff}$) has been discussed in
$\cite{voloshin3}$. With $\omega \equiv \sqrt{{\bf P}^{2}+1}$, the Green
function for the operator in equation ($\ref{xformedGenDiff}$) was found to be
\beq
G_{\omega}(t,t') = {f^{-}_{\omega}(t)f^{+}_{\omega}(t')\theta(t-t')
+ f^{-}_{\omega}(t')f^{+}_{\omega}(t)\theta(t'-t)\over{2\omega}},
\label{GreenDef}
\eeq
where
\beq
f^{\pm}_{\omega}(t) = e^{\mp\omega t}
F(-2,3;1\pm\omega;{1\over{1+e^{2t}}}),
\eeq
and $F(a,b;c;x)$ is the hypergeometric function.

With the aid of this Green function, the solution of equation
$(\ref{xformedGenDiff})$ is found to be,
\beq
g^{(N)}(t) = \lambda e^{-\epsilon t} \int_{-\infty}^{+\infty}
{G_{\omega}(t,t')\sum_{N_{1},N_{2},N_{3}<N} \delta_{N,N_1+N_2+N_3}
y^{N_1}(t') y^{N_2}(t') y^{N_3}(t') dt'}.
\label{gensol}
\eeq

Equation ($\ref{gensol}$) allows one to
recursively calculate any $g^{(N)}$ if one knows all $g^{(M)}$ for all
$M<N$. Equation ($\ref{gensol}$) can be summarized by a set of Feynman
rules for calculating amplitudes with multiparticle production in the
background.

1. Draw all possible diagrams with the external legs labelled. Do not
include any propagators that represent particles at rest.

2. Assign a unique time label to every vertice and to the endpoint of
every external leg.

3. For every external leg, multiply the amplitude by a factor of
$e^{\epsilon t}$, where $\epsilon$ is the energy of the of the particle
at the endpoint of the leg, and $t$ is the time label at the endpoint of
the leg.

4. For every line, multiply by a factor of $G_{\omega}(t_1,t_2)$ where
$t_1$ and $t_2$ are the time labels of the two endpoints of the line, and
$\omega$ is the expected on-shell energy of the momenta flowing through the
line, i.e. $\omega = \sqrt{{\bf k}^2 +1}$.

5. For every vertice, multiply by a factor if $\lambda$. In addition,
for every three-point vertice multiply by a factor of $g^{(0)}(t)$. This
last factor is present to account for the production of particles at the
vertice.

6. Integrate over all time labels except for one.

7. Multiply by a factor of $e^{-\sum \epsilon_i t}$, where $t$ is the
time label that was not integrated over and the sum includes the label at
every external leg.

8. Each diagram must be multiplied by an appropriate symmetry factor.

Note that spatial momenta is conserved at
every vertice and over all. Energy, however, is not conserved. This basic
procedure can be used to construct Feynman rules in a number of theories
including $\lambda\phi^4$ in both the case of broken and unbroken symmetry.

For the case of unbroken symmetry, the Green function for the propagator
is given by equation ($\ref{GreenDef}$), and the vertice amplitude is given
by equation ($\ref{gdef}$). The rules will generate $g^{(N)}(t)$. To find
the generating function of the amplitudes, one simply has to make the
substitution,
\beq
z(t) \equiv i \sqrt{8\over{\lambda}} e^{t}.
\eeq
The amplitude is given by the $n^{th}$ derivative of $g^{(N)}(z)$,
\beq
A^{(N)}_{n}(p_1,...,p_N) = {d^{n}\over{dz^{n}}} g^{(N)}(p_1,...,p_N;z).
\eeq

In theories with a broken symmetry, the propagator is given
by$^{\cite{smith}}$,
\beq
G_{\omega}(t,t') = {f^{-}_{\omega}(t)f^{+}_{\omega}(t')\theta(t-t')
+ f^{-}_{\omega}(t')f^{+}_{\omega}(t)\theta(t'-t)\over{2\omega}},
\label{ubGreenDef}
\eeq
where
\beq
f^{\pm}_{\omega}(t) = e^{\mp\omega t}
F(-2,3;1\pm 2\omega;{1\over{1+e^t}}).
\eeq
The generating function for the amplitude of one particle production is
given by, \beq
g^{(0)}(t) = \sqrt{1\over{2\lambda}} \left ( {1-e^{t}\over{1+e^{t}}} \right
). \eeq
The amplitudes are found in the same manner as in the case of unbroken
symmetry. However, in the case of broken symmetry, one must make the
substitution,
\beq
z(t) \equiv -\sqrt{2\over{\lambda}} e^{t}.
\label{bzdef}
\eeq
Rules for other theories may be constructed if one can calculate the
propagator and the generating function for $1 \to n$ processes.

As an example, we shall calculate the tree-level amplitude for threshold
production  of $n$ particle by three on-shell particles in a theory with
spontaneously broken symmetry. This amplitude is divergent when the
incoming particles are on-mass-shell. To consider the amplitude of the real
process, one must take the residue of the triple pole as the energy of all
three incoming particles tends to its mass-shell value. The generating
functional can be seen from the diagram in figure 2  to be,
\beq
g^{(3)}(t) = 6\lambda \int e^{\epsilon_1 t_1}G_{\omega_1}(t,t_1)
e^{\epsilon_2 t_2}G_{\omega_2}(t,t_2)
e^{\epsilon_3 t_3}G_{\omega_1}(t,t_3)
 \sqrt{1\over{2\lambda}} \left ( {1-e^{t}\over{1+e^{t}}} \right )
dt_{1}dt_{2}dt_{3}.
\label{3ptexp}
\eeq

Taking the on-shell limit corresponds to the limit $\epsilon_{i} \to
\omega_{i}$. Inspection of equation ($\ref{ubGreenDef}$) reveals that in
order to have poles in the amplitude, the time ordering of the three
propagators is uniquely determined, $t_{1},t_{2},t_{3} < t$. Causality
considerations would also lead one to believe that this is the only time
ordering capable of producing a physical amplitude.

Equation ($\ref{3ptexp}$) contains three integrals over hypergeometric
functions. These are easiest to handle by writing the hypergeometric
function in terms of the standard hypergeometric series. The integral of
interest in equation ($\ref{3ptexp}$) is
\beq
\int_{-\infty}^{t}
e^{(\omega+\epsilon)t'} F(-2,3;1-2\omega;{1\over{1+e^{t'}}}) dt' =
\int_{-\infty}^{t}
e^{(\omega+\epsilon)t'} \sum_{i=1}^{3} {(-2)_{n}
(3)_{n}\over{(1+2\omega)_{n}}} {1\over{1+e^{t'}}} dt',
\label{basInt}
\eeq
where $(a)_{n}$ is the Pochhammer symbol,
\beq
(a)_{n} \equiv {\Gamma(a+n+1)\over{\Gamma(a+1)}}.
\eeq

When the integrand in equation ($\ref{basInt}$) is expanded in a series in
$e^{t'}$, it is easy to see that only the first term in the series shall
contribute to the pole as $\epsilon \to -\omega$. The residue of the pole
in equation ($\ref{basInt}$) is the simple expression
$F(-2,3;1-2\omega;1)$. We can now recast the integrals in equation
($\ref{3ptexp}$) in the simple form,
\beq
\lim_{\epsilon \to -\omega} (\epsilon^{2}-\omega^{2})
\int_{-\infty}^{\infty} e^{\epsilon t'} G_{\omega}(t,t') dt' =
{1\over{(1+e^{t})^2}}F(-2,-2-2\omega;1-2\omega;-e^{t}).
\label{intResult}
\eeq

Using the definition of $z$ in equation ($\ref{bzdef}$), we can
write an expression for the amplitude in equation ($\ref{3ptexp}$),
\begin{eqnarray}
g^{(3)}(z) = 6 \sqrt{\lambda\over{2}}
{(1+\sqrt{\lambda/2}z)\over{(1-\sqrt{\lambda/2}z)^7}}
(1-4\sqrt{\lambda\over{2}}{(1+\omega_{1})\over{1+2\omega_{1}}}z+
{\lambda(1+\omega_{1})\over{2\omega_{1}}}z^{2})
\nonumber \\
(1-4\sqrt{\lambda\over{2}}{(1+\omega_{2})\over{1+2\omega_{2}}}z+
{\lambda(1+\omega_{2})\over{2\omega_{2}}}z^{2})
(1-4\sqrt{\lambda\over{2}}{(1+\omega_{3})\over{1+2\omega_{3}}}z+
{\lambda(1+\omega_{3})\over{2\omega_{3}}}z^{2})
\label{3ptans}
\end{eqnarray}

Equation ($\ref{3ptans}$) is the generating function for the tree-level
amplitudes of $3 \to n$ processes at the kinematical threshold. In
the calculation of an amplitude, $\omega_{1}$, $\omega_{2}$,
$\omega_{3}$, and the number of particles are not four independent
parameters. These four parameters can be reduced to two independent
parameters through the application of momenta and energy conservation
principles.

The $3 \to n$ amplitudes contained in the generating function in equation
($\ref{3ptans}$) can be written in a compact form in the asymptotic
limit, i.e. when $\omega_1, \omega_2, \omega_{3}$, and $n >> 1$. The
tree-level amplitude for $3 \to n$ production at the kinematic threshold
in the large $n$ approximation in the theory with broken symmetry is
\beq
A^{0}(3 \to n) = {2\over{5!}} \left (
{\lambda\over{2}} \right )^{(n+1)/2} {n! n^{6}\over{(2\omega_{1})^{2}
(2\omega_{2})^{2}(2\omega_{3})^{2}}}.
\label{3ptamp}
\eeq

Voloshin has noted the $2 \to n$ at the one-loop level can be related to
the tree-level $3 \to n$ amplitude$^{\cite{voloshin6}}$. This reasoning
is based on the argument that in the spontaneously broken theory, the
imaginary part of $1 \to n$ processes must vanish to all orders in
perturbation theory. One can consider the unitary cuts on the one-loop $1
\to n$ amplitude to derive the condition,
\begin{eqnarray}
2 Im A^{1}(1 \to n) = \int A^{1}(1 \to 2)A^{0}(2 \to n) d\tau_{2} +
\nonumber \\
\int A^{0}(1 \to 2)A^{1}(2 \to n) d\tau_{2} +
\int A^{0}(1 \to 3)A^{0}(3 \to n) d\tau_{3},
\label{unitcond}
\end{eqnarray}
where the integration is done over the phase space of the intermediate
particles, and the superscripts denote the number of loops in the
amplitude. The first term on the right hand side of equation
($\ref{unitcond}$) is zero due to the vanishing of $A^{0}(2 \to n)$ for
all $n>2$ as noted in $\cite{smith}$. Since the left hand side of equation
(\ref{unitcond}) is zero, the one-loop $2 \to n$ amplitude may be related
to the tree-level $3 \to n$ amplitude in equation ($\ref{3ptans}$). The
integration over the phase space of equation ($\ref{3ptans}$) is
non-trivial, but can determine whether the $2 \to n$ nullification noted
in $\cite{smith}$ will continue at the one-loop level.

The expression in equation ($\ref{3ptamp}$) can now be combined with the
$1 \to 3$ amplitude to find the contribution of the second term in the
unitary relation, equation ($\ref{unitcond}$). This contribution is of
the order of $\lambda^{(n+1)/2}n!n^2$. The order of magnitude of the
one-loop correction to the $2 \to n$ amplitude is, therefore,
\beq
A^{1}(2 \to n) = \lambda^{(n+2)/2} n! n^{2}.
\label{2ton}
\eeq
Although the tree-level contribution vanishes in $\lambda\phi^4$ theory,
the one-loop contribution can be compared with the tree-level amplitude
for $2 \chi \to n \phi$ in an interactive theory as calculated by Brown
and Zhai$^{\cite{bz}}$. They found that the tree level $2 \to n$
amplitude should grow like $\lambda^{n/2}n! n$.

The above arguments indicate that although the  $2 \to n$
tree-level threshold amplitude may vanish, diagrams at the one-loop level
provide a correction of order $\lambda n$. If the loop expansion were a
series expansion in powers of $\lambda n^2$, the results of ${\cite{bz}}$
would indicate that the one-loop $2 \to n$ correction should grow like
$\lambda^{(n+1)/2}n! n^3$. There is a $1/n$ suppression in the one-loop
correction to threshold $2 \to n$ processes.

 The rules discussed in this paper provide a compact method for
calculating many processes involving large numbers of particles at the
kinematic threshold. They provide a useful tool for investigating
suppression of amplitudes in multiparticle processes. They may prove
to be of further assistance in studying
multiparticle phenomena such as the vanishing of $2 \to n$ amplitudes.

I would like to thank M.B. Voloshin for several helpful discussions.

%\end{document}
\newpage
\thicklines
\unitlength=1.00mm
%\special{em:linewidth 0.4pt}
%\linethickness{0.4pt}
\begin{picture}(154.00,117.00)(0,40)
\put(39.00,93.00){\circle*{5.20}}
\put(36.00,93.00){\line(-1,0){15.00}}
\put(42.00,93.00){\line(1,0){16.00}}
\put(37.00,95.00){\line(-2,3){6.00}}
\put(38.00,96.00){\line(-1,6){1.33}}
\put(40.00,96.00){\line(0,1){9.00}}
\put(41.00,95.00){\line(2,5){4.00}}
\put(41.00,91.00){\line(6,-5){12.00}}
\put(41.00,92.00){\line(3,-1){14.00}}
\put(42.00,94.00){\line(3,1){14.00}}
\bezier{100}(29.00,107.00)(38.00,116.00)(46.00,107.00)
\put(38.00,116.00){\makebox(0,0)[cc]{$n$}}
\put(61.00,99.00){\makebox(0,0)[cc]{$p_1$}}
\put(61.00,93.00){\makebox(0,0)[cc]{$p_2$}}
\put(60.00,88.00){\makebox(0,0)[cc]{$p_3$}}
\put(56.00,79.00){\makebox(0,0)[cc]{$p_n$}}
\put(59.00,84.00){\circle{0.00}}
\put(58.25,82.50){\circle{0.00}}
\put(57.50,81.00){\circle{0.00}}
\put(77.00,93.00){\makebox(0,0)[cc]{=}}
\put(85.00,93.00){\makebox(0,0)[cc]{{\huge $\Sigma$}}}
\put(96.00,93.00){\line(1,0){23.00}}
\put(119.00,109.00){\circle*{5.20}}
\put(119.00,72.00){\circle*{5.20}}
\put(140.00,93.00){\circle*{5.20}}
\put(119.00,93.00){\line(0,1){13.00}}
\put(119.00,93.00){\line(0,-1){18.00}}
\put(119.00,93.00){\line(1,0){18.00}}
\put(122.00,110.00){\line(5,3){10.00}}
\put(122.00,109.00){\line(1,0){10.00}}
\put(121.00,107.00){\line(2,-1){11.00}}
\put(122.00,72.00){\line(1,0){10.00}}
\put(121.00,73.00){\line(2,1){11.00}}
\put(121.00,70.00){\line(2,-1){11.00}}
\put(116.00,72.00){\line(-1,0){6.00}}
\put(116.00,73.00){\line(-5,3){5.00}}
\put(117.00,70.00){\line(-3,-1){6.00}}
\put(116.00,109.00){\line(-1,0){6.00}}
\put(117.00,110.00){\line(-3,2){6.00}}
\put(117.00,108.00){\line(-3,-2){6.00}}
\bezier{72}(109.00,115.00)(102.00,109.00)(109.00,103.00)
\bezier{76}(109.00,77.00)(101.00,72.00)(109.00,66.00)
\bezier{108}(134.00,80.00)(144.00,74.00)(134.00,62.00)
\bezier{84}(134.00,117.00)(140.00,109.00)(134.00,100.00)
\put(141.00,96.00){\line(2,3){4.00}}
\put(140.00,96.00){\line(1,6){1.00}}
\put(142.00,94.00){\line(1,1){5.00}}
\put(143.00,92.00){\line(3,-2){9.00}}
\put(141.00,91.00){\line(1,-1){8.00}}
\put(140.00,90.00){\line(2,-5){3.67}}
\bezier{68}(141.00,104.00)(149.00,107.00)(149.00,99.00)
\bezier{76}(154.00,86.00)(154.00,77.00)(144.00,79.00)
\put(102.00,108.00){\makebox(0,0)[cc]{$n_1$}}
\put(142.00,110.00){\makebox(0,0)[cc]{$m_1$}}
\put(102.00,71.00){\makebox(0,0)[cc]{$n_3$}}
\put(144.00,71.00){\makebox(0,0)[cc]{$m_3$}}
\put(148.00,107.00){\makebox(0,0)[cc]{$n_2$}}
\put(151.00,76.00){\makebox(0,0)[cc]{$m_2$}}
\put(80.00,32.00){\makebox(0,0)[cc]{{\bf Figure 1} - Symbolic
representation of recursion relation}} \end{picture}

\newpage

\begin{picture}(83.00,121.00)(0,40)
\bezier{392}(40.00,116.00)(60.00,71.00)(80.00,116.00)
\put(60.00,94.00){\line(0,1){22.00}}
\put(44.00,116.00){\makebox(0,0)[cc]{$t_1$}}
\put(63.00,116.00){\makebox(0,0)[cc]{$t_2$}}
\put(83.00,116.00){\makebox(0,0)[cc]{$t_3$}}
\put(60.00,89.00){\makebox(0,0)[cc]{$t$}}
\put(40.00,121.00){\makebox(0,0)[cc]{$p_1$}}
\put(60.00,121.00){\makebox(0,0)[cc]{$p_2$}}
\put(80.00,121.00){\makebox(0,0)[cc]{$p_3$}}
\put(60.00,58.00){\makebox(0,0)[cc]{{\bf Figure 2} - Feynman diagram
for $3 \to n$ processes}} \end{picture}

\end{document}